\documentclass[pra,notitlepage,twocolumn]{revtex4-1}

\usepackage[T1]{fontenc}
\usepackage{amsmath,amssymb,amsfonts,amsthm,diagbox,dsfont,xcolor,graphicx}
\usepackage[colorlinks,breaklinks]{hyperref}
\definecolor{darkred}{rgb}{0.5,0,0}
\definecolor{darkgreen}{rgb}{0,0.5,0}
\definecolor{darkblue}{rgb}{0,0,0.5}
\hypersetup{linkcolor=darkred,citecolor=darkgreen,urlcolor=darkblue}

\def\ket#1{|#1\rangle}
\def\braket#1#2{\langle#1|#2\rangle}
\def\ketbra#1{|#1\rangle\langle#1|}
\def\bra#1{\langle#1|}

\newcommand{\irg}{\eta^\mathrm{g}}
\newcommand{\IRG}{H^\mathrm{g}}
\newcommand{\id}{\mathds{1}}
\newcommand{\vj}{{\vec{j}}}
\newcommand{\IR}{\mathrm{IR}}
\newcommand{\SR}{\mathrm{SR}}

\renewcommand{\leq}{\leqslant}
\renewcommand{\geq}{\geqslant}
\newtheorem{definition}{Definition}
\DeclareMathOperator{\Tr}{Tr}

\begin{document}

\title{Robust genuine high-dimensional steering with many measurements}
\author{S{\'e}bastien Designolle}
\affiliation{Department of Applied Physics, University of Geneva, 1211 Geneva, Switzerland}
\date{18 March 2022}

\begin{abstract}
  Quantum systems of high dimensions are attracting a lot of attention because they feature interesting properties when it comes to observing entanglement or other forms of correlations.
  In particular, their improved resistance to noise is favourable for experiments in quantum communication or quantum cryptography.
  However, witnessing this high-dimensional nature remains challenging, especially when the assumptions on the parties involved are weak, typically when one of them is considered as a black box.
  In this context, the concept of genuine high-dimensional steering has been recently introduced and experimentally demonstrated [Phys.~Rev.~Lett.~126, 200404 (2021)]; it allows for a one-sided device-independent certification of the dimension of a bipartite shared state by only using two measurements.
  Here I overcome this limitation by developing, for more than two measurements, universal bounds on the incompatibility robustness, turned into meaningful dimension certificates.
  Interestingly, even though the resulting bounds are quite loose, they still often offer an increased resistance to noise and could then be advantageously employed in experiments.
\end{abstract}

\maketitle

\section{Introduction}
\label{sec:introduction}

Quantum physics allows for correlations that do not have any classical explanation.
The typical scenario used to witness these quantum correlations involves two parties performing measurements on their half of a bipartite shared state whose quantumness can eventually lead to the demonstration of, for instance, Bell nonlocality~\cite{BCP+14} or Einstein--Podolsky--Rosen steering~\cite{UCNG20}.
The former soon made its way to the laboratory while the latter, formalised quite recently~\cite{WJD07}, received more experimental attention during the last decade.

Both phenomena can leverage high- or infinite-dimensional shared states to feature a better resistance to noise and losses, thus becoming more resistant to attacks~\cite{HES+12,HRAR15,SC18,GCH+19,WXK+20,DLW+21}.
In turn, they can also allow for a certification of the dimension of the underlying state, with no characterisation of the devices for nonlocality, and only on one side for steering.
Crucially, in a cryptographic context, using a high-dimensional state in practice does not guarantee that the resulting correlations cannot be theoretically obtained by a lower-dimensional resource, so that a few works devoted some effort to find suitable witnesses~\cite{BPA+08,WYBS17,DSU+21}.

In this context, the concept of genuine high-dimensional steering has been recently defined~\cite{DSU+21} and typically allows for a one-sided device-independent certificate of the dimension of the shared state, through the formal concept of Schmidt number defined below.
However, the reach of the theory in Ref.~\cite{DSU+21} is quite limited because only pairs of measurements are discussed, while it can often be desirable to investigate more, in particular to increase the robustness to noise (a caricature being Ref.~\cite{Ver08} and its 465 measurements).

In this paper I go beyond this limitation by deriving dimension witnesses for more than two measurements.
The steps followed are the following: unfold the proof for pairs, extract its key ingredient, use it repeatedly to extend it for more measurements, and wrap the proof back up to obtain certificates.
Perhaps surprisingly, this seemingly simple procedure nonetheless gives rise to nontrivial bounds, which are, however, not tight.
Moreover, their noise tolerance, that is, the affordable imperfection of the shared state, often improves on the one for pairs.
Should the bounds have been tight, this last feature would have been completely expected, but, since this is not the case, it indicates their quality, or at least their usefulness.

The article is organised as follows: after recalling the main definitions needed for genuine high-dimensional steering (Secs~\ref{sec:preliminaries} and \ref{sec:HDS}), I draw the known quantitative link between the problem of obtaining a dimension certificate and the one of finding dimension-dependent bounds on measurement incompatibility (Sec.~\ref{sec:incompatibility}).
Then I derive the main theoretical results of this article, dealing with the latter problem (Sec.~\ref{sec:universal}), before importing them back to the former (Sec.~\ref{sec:certification}).
Finally, I use the concrete example of projective measurements onto mutually unbiased bases to show how to take advantage of the new witnesses, which often exhibit an improved robustness to noise (Sec.~\ref{sec:example}).
Note that the impatient reader only willing to experimentally apply the results may skip Secs~\ref{sec:incompatibility} and \ref{sec:universal}.

\section{Preliminaries}
\label{sec:preliminaries}

In this section I introduce the different notations used throughout this article and briefly recall the definition of steering.

\begin{figure}[h]
  \centering
  \includegraphics[width=\columnwidth]{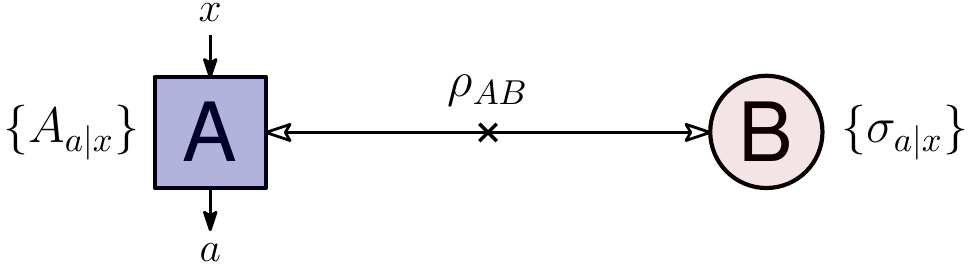}
  \caption{
    Steering scenario: Alice applies the measurements $\{A_{a|x}\}$, steering Bob's system to the corresponding conditional states $\{\sigma_{a|x}\}$.
  }
  \label{fig:steering}
\end{figure}

Let me start with the usual steering scenario~\cite{CS16b,UCNG20}.
Alice and Bob share an entangled quantum state $\rho_{AB}$.
Let Bob be allowed to fully characterise his part of the shared state, while Alice performs the measurements $\{\{A_{a|x}\}_a\}_x$, simply denoted $\{A_{a|x}\}$ or $A$ in the following, see Fig.~\ref{fig:steering}.
Loosely speaking, this corresponds to a situation in which Bob is trusted, but not Alice; this is why I will sometimes refer to this scenario as \emph{one-sided device-independent}.
Note that by \emph{measurement} I always mean a positive operator-valued measure (POVM), that is, a set of positive operators summing up to identity.
In the rest of this article, the measurement outcome $a$ will range from 1 to $d$ while the setting $x$ will be between 1 and $k$, the number of measurements.

Once Alice has performed her measurement, Bob is left with a subnormalised assemblage of states
\begin{equation}
  \label{eqn:assemblage}
  \sigma_{a|x}:=\Tr_A\big[(A_{a|x}\otimes \id_B)\rho_{AB}\big],
\end{equation}
with the condition ${\sum_a\sigma_{a|x}=\rho_B:=\Tr_A (\rho_{AB})}$ for all $x$ corresponding to no-signalling.

To define the notion of unsteerability, I need the concept of local hidden state (LHS) model.
This refers to a situation in which a classical message $\lambda$ is sent to Alice while a quantum state $\sigma_\lambda$ is sent to Bob.
Then Alice can decide her output $a$ with a probability $p_A(a|x,\lambda)$, based on her input $x$ and the variable $\lambda$, distributed with density $\pi(\lambda)$.
The resulting assemblage, formally written in Eq.~\eqref{eqn:LHS} below, is called a LHS model and its (in)existence is at the heart of the concept of (un)steerability.

\begin{definition}
  An assemblage $\{\sigma_{a|x}\}$ \emph{demonstrates steering} from Alice to Bob if it admits no LHS model, that is, no decomposition of the form
  \begin{equation}
    \label{eqn:LHS}
    \sigma_{a|x}=\int p_A(a|x,\lambda)\pi(\lambda)\sigma_\lambda\,\mathrm{d}\lambda,
  \end{equation}
  for a variable $\lambda$ distributed with density $\pi(\lambda)$ and a local response distribution $p_A (a|x,\lambda)$.
  \label{def:steering}
\end{definition}

At the level of the shared state $\rho_{AB}$, I say that $\rho_{AB}$ is \emph{steerable} from Alice to Bob if there exist measurements $\{A_{a|x}\}$ such that the resulting assemblage~\eqref{eqn:assemblage} demonstrates steering.

The notion of steerability just defined naturally fits in between those of entanglement~\cite{GT09} and nonlocality~\cite{BCP+14}, as can be foreseen from the scenario: an entanglement test would involve two parties able to fully characterise their local state, while a Bell test would have both of them taking inputs and giving outputs.

\section{Genuine high-dimensional steering}
\label{sec:HDS}

In this section I recall the concept defined in Ref.~\cite{DSU+21}.
The idea is that an assemblage exhibits genuine $n$-dimensional steering when it can provably not have been prepared by using states with dimension at most $n-1$.
Importantly, this notion ignores the actual way this assemblage has been produced and focuses on all possible preparations.
There is indeed no guarantee that the preparation implemented is optimal and it is very often the case that there is, at least theoretically, a better way using lower-dimensional states.
As will be then seen in Sec.~\ref{sec:certification}, this concept naturally leads to the possibility of certifying the dimension in a one-sided device-independent manner.

The first step is to formalise the above-mentioned notion of ``all possible preparations.''
It boils down to the one of Schmidt number through the following definitions.

\begin{definition}
  The \emph{Schmidt rank} of a bipartite pure state is the number of nonzero terms in its Schmidt decomposition.
  \label{def:SR}
\end{definition}

The Schmidt decomposition is a standard and fundamental tool in quantum information.
The above definition for pure states can then be extended to mixed states as follows.

\begin{definition}
  A bipartite state $\rho_{AB}$ has \emph{Schmidt number} $n$ when it can be written as ${\rho_{AB}=\sum_i p_i \ket{\varphi_i}\bra{\varphi_i}}$, where the bipartite pure states $\ket{\varphi_i}$ have Schmidt rank at most $n$ for all $i$.
  \label{def:SN}
\end{definition}

This naturally coincides with the Schmidt rank for pure states.
Note that, in general, the number of possible decomposition is infinite.
In this article, I sometimes refer to the Schmidt number simply as the `dimension'.

\begin{definition}
  An assemblage $\{\sigma_{a|x}\}$ acting on $\mathds{C}^d$ is \emph{$n$-preparable} (with ${1\leq n\leq d}$) when it can be written as ${\sigma_{a|x} = \Tr_A[ (A_{a|x} \otimes \id_B) \rho_{AB}]}$ where the bipartite state $\rho_{AB}$ has Schmidt number at most $n$.
  \label{def:HDS}
\end{definition}

\begin{figure}[h]
  \centering
  \includegraphics[width=0.8\columnwidth]{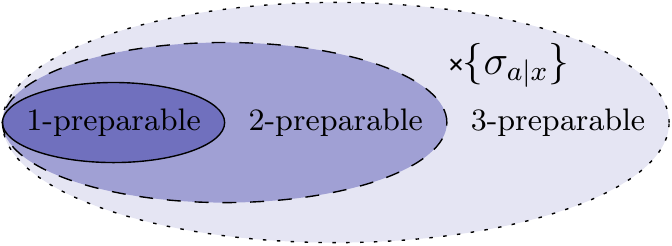}
  \caption{
    High-dimensional steering: the assemblage $\{\sigma_{a|x}\}$ is 3-preparable but not 2-preparable, it exhibits genuine three-dimensional steering.
    Certifying this property would attest that the underlying state must have a Schmidt number of (at least) three.
  }
  \label{fig:dolls}
\end{figure}

The set of all $n$-preparable assemblages is convex and a subset of the one of all $(n+1)$-preparable assemblages, so that a Russian doll structure emerges, see Fig.~\ref{fig:dolls}.
The property of 1-preparability is equivalent to the existence of a LHS model.
Then I say that an assemblage exhibits genuine $n$-dimensional steering when it is $n$-preparable but not $(n-1)$-preparable~\cite{DSU+21}.

This property is hard to characterise, particularly because it is unknown whether it can be solved by means of a semidefinite program (SDP)~\cite{BV04} as is the case for unsteerability~\cite{CS16b}.
However, there exist experimental-friendly witnesses able to detect it; for instance, when Alice is allowed to perform two measurements only, Ref.~\cite{DSU+21} provides a tight criterion.
In Secs~\ref{sec:incompatibility} and \ref{sec:universal}, I recall how it was obtained and generalise it to more than two measurements.
The reader only interested in using the resulting certificates can directly jump to Sec.~\ref{sec:certification} where all relevant information is provided.

\section{Connection with incompatibility}
\label{sec:incompatibility}

In this section, I briefly recall the link between steering and incompatibility which is at the heart of the method used in Ref.~\cite{DSU+21} to assess the presence of genuine high-dimensional steering.
The idea can be traced back to Refs.~\cite{QVB14,UMG14,UBGP15}: mathematically, the existence of a LHS is very similar to the one of a parent measurement, the central concept in the theory of incompatibility.

More formally, by denoting $\delta$ the Kronecker delta and $\vj:=j_1\ldots j_k$, where $k$ is the number of measurements involved, I have the following definition.

\begin{definition}
  The measurements $\{A_{a|x}\}$ are \emph{incompatible} when they do not admit a parent measurement, that is, a measurement $\{G_\vj\}$ such that
  \begin{equation}
    \sum\limits_\vj \delta_{j_x,a} G_\vj = A_{a|x}
    \label{eqn:parent}
  \end{equation}
  for all $a$ and $x$.
  \label{def:incompatibility}
\end{definition}

The analogy between Eqs.~\eqref{eqn:LHS} and \eqref{eqn:parent} comes from the fact that the deterministic strategy can be chosen to be simple marginals, see, for instance, Ref.~\cite{UBGP15}.
I will not detail this point but rather move on to the quantitative link that can be drawn between steering and incompatibility quantifiers for $n$-preparable assemblages.

\begin{definition}
  The \emph{steering robustness} of an assemblage $\{\sigma_{a|x}\}$ is
  \begin{equation}\label{eqn:SR}
    \!\!\!\SR_{\{\sigma_{a|x}\}}:=\!\min_{t,\{\tau_{a|x}\}}\Bigg\{t\geq 0\,\bigg\vert\, \frac{\sigma_{a|x} + t{\tau_{a|x}}}{1+t}\text{ unsteerable}\Bigg\},
  \end{equation}
  where $\{\tau_{a|x}\}$ is a valid assemblage.
  \label{def:sr}
\end{definition}

I explain how to experimentally estimate the steering robustness in Sec.~\ref{sec:sr}.
This quantity is obviously zero for unsteerable assemblages.
It measures the tolerance of a steerable assemblage subject to noise~\cite{PW15}.
Importantly, it can be cast as a SDP~\cite{CS16b} which makes it quite tractable.
A very similar quantifier can be defined for incompatibility, the incompatibility robustness, denoted by $\IR_{\{A_{a|x}\}}$~\cite{CS16a}.
For consistency with Ref.~\cite{DFK19} as well as for a better understanding of the mechanism employed below in Sec.~\ref{sec:universal}, I use here a different parametrisation, strictly equivalent to $\IR$ as will be seen in Eq.~\eqref{eqn:IR_irg}.

\begin{definition}
  The \emph{incompatibility generalised robustness} of the measurements $\{A_{a|x}\}$ is the solution of the SDP
  \begin{align}
    \irg_{\{A_{a|x}\}}&:=\left\{
      \begin{array}{cl}
        \max\limits_{\eta,\{G_\vj\}} &\eta \\
        \mathrm{s.t.}
        &G_\vj \geq 0,\quad\sum\limits_\vj G_\vj = \id, \\
        &\sum\limits_\vj \delta_{j_x,a} G_\vj \geq \eta A_{a|x}.
      \end{array}
    \right.\label{eqn:irg}
  \end{align}
\end{definition}

While $\IR$ is 0 for compatible measurements, $\irg$ is 1 for them.
Moreover, the more incompatible the measurements are, the bigger $\IR$ but the smaller $\irg$.
In general, the following equality holds:
\begin{equation}
  \IR_{\{M_{a|x}\}}=\frac{1}{\irg_{\{M_{a|x}\}}}-1.
  \label{eqn:IR_irg}
\end{equation}

With these tools at hand, I can state the main claim of Ref.~\cite{DSU+21}.
With Def.~\ref{def:HDS}, the convexity of $\SR$, and standard relations between quantifiers, it is proven therein that an $n$-preparable assemblage $\{\sigma_{a|x}\}$ must satisfy
\begin{equation}
  \SR_{\{\sigma_{a|x}\}}\leq\max_{\{M_{a|x}\}}\IR_{\{M_{a|x}\}},
  \label{eqn:SRup}
\end{equation}
where the optimisation is performed over sets of $k$ measurements of dimension at most $n$.
With Eq.~\eqref{eqn:IR_irg} in mind I introduce
\begin{equation}
  \IRG_{k,n}:=\min\limits_{\{M_{a|x}\}}\irg_{\{M_{a|x}\}},
  \label{eqn:IRG}
\end{equation}
so that Eq.~\eqref{eqn:SRup} reads, for $n$-preparable assemblages,
\begin{equation}
  \SR_{\{\sigma_{a|x}\}}\leq\frac{1}{\IRG_{k,n}}-1.
  \label{eqn:SRup2}
\end{equation}

For pairs of measurements, i.e., when $k=2$, the main result of Ref.~\cite{DFK19} on most incompatible pairs of measurements applies and gives
\begin{equation}
  \IRG_{2,n}=\frac12\left(1+\frac{1}{\sqrt{n}}\right).
  \label{eqn:IRG2}
\end{equation}
I derive again this result below in Sec.~\ref{sec:pairs}.

As a summary of this section, I have presented here a connection that allows to reformulate the problem of finding necessary conditions for $n$-preparability of assemblages in terms of an incompatibility problem: finding bounds on the incompatibility robustness attainable by measurements in dimension $n$.
In Sec.~\ref{sec:universal} below, I recall known facts about this problem and eventually give new results going beyond pairs of measurements.

\section{Universal incompatibility bounds}
\label{sec:universal}

In this section, I recall the cloning machine technique used to get a very general bound on incompatibility.
Afterwards I briefly derive the tight bound obtained in Ref.~\cite{DFK19} for pairs of measurements.
Elaborating on this I can then state the main results of this article, namely, loose bounds for more that two measurements.

\subsection{Cloning machine bound}
\label{sec:cloning}

To perform several measurements simultaneously, a natural idea suggested by classical intuition is to duplicate the input state and then to feed each measurement with one of the copies.
However, in quantum mechanics, by virtue of the no-cloning theorem, the duplication process cannot be perfect, so that the resulting measurements are noisy.
This process is nonetheless a good starting point to find general bounds on the incompatibility robustness.

For the incompatibility depolarising robustness, a measure of incompatibility different from $\irg$, a cloning machine bound is, for instance, presented in Ref.~\cite[Eq.~(11)]{HMZ16}, and can then be converted to a bound on $\irg$ thanks to the relation between the two measures described in Ref.~\cite[Appendix~B]{DFK19}, giving rise to
\begin{equation}
  \IRG_{k,n} \geq \frac1k\left(1+2\ \frac{k-1}{n+1}\right).
  \label{eqn:irg_cloning}
\end{equation}
The proof relies on the symmetric cloning machine from Ref.~\cite{KW99}.

\subsection{Tight bound for pairs of measurements}
\label{sec:pairs}

Following Ref.~\cite{DFK19}, for a pair $\{A,B\}$ of rank-one measurements in dimension $n$, I introduce a feasible point $\{G_{ab}\}$ for the optimisation in Eq.~\eqref{eqn:irg}, that is,
\begin{equation}
  \label{eqn:irg_low_ansatz}
  \begin{aligned}
    G_{ab}\propto\,&\big\{A_a,B_b\big\} + \frac{1}{2\sqrt{n}}\big[\Tr(B_b)A_a + \Tr(A_a)B_b\big]\\
    &+ \frac{\sqrt{n}}{2}\left(A_a^\frac12 B_b A_a^\frac12 + B_b^\frac12 A_a B_b^\frac12\right).
  \end{aligned}
\end{equation}
Thanks to the operator inequality
\begin{equation}
  \sum_bB_b^{\frac12} A_a B_b^{\frac12} \geq \frac1n A_a,
\end{equation}
which is ultimately a consequence of the Cauchy--Schwarz inequality, the parent measurement from Eq.~\eqref{eqn:irg_low_ansatz} gives the bound
\begin{equation}
  \label{eqn:irg_low}
  \frac12\left(1+\frac{1}{\sqrt{n}}\right)\leq\irg_{\{A,B\}}.
\end{equation}
Due to the properties of the measure $\irg$, this can be extended to any pair of measurements (not necessarily rank-one), see Ref.~\cite{DFK19}.
This concludes the proof of Eq.~\eqref{eqn:IRG2}.

Note that it is also proven in Ref.~\cite{DFK19} that pairs of projective measurements onto mutually unbiased bases, a notion discussed in Sec.~\ref{sec:MUB}, do actually reach the bound~\eqref{eqn:irg_low} for all dimensions $n$.
However, these pairs of measurements are not the only ones to saturate this bound; see the recent article~\cite{MK21} for examples and related discussions.

\subsection{Loose bounds for more measurements}
\label{sec:more}

When considering $k>2$, that is, strictly more than two measurements, the problem scales up in complexity.
There are, for instance, examples of three measurements that are pairwise compatible but triplewise incompatible~\cite{Bus86}.
Even though the above bounds do only capture the pairwise interaction among those larger sets of measurements, I investigate here how they can be useful to get nontrivial bounds in this general case.

The idea is to successively apply the universal lower bounds for pairs of measurements.
Such a procedure is schematically described in Ref.~\cite[Appendix~E.2]{DFK19} but I could find another one since then, which is always better (often strictly) and which I present in detail here.
The idea is to always go back to the case of a number of measurements being a power of two, as in this case, it is obvious to see that applying repeatedly the lower bound for pairs of measurements will lead to a general lower bound being a suitable power of it.
Specifically, if $k=2^r+l$ with $0\leq l<2^r$, then I first pair $2l$ measurements using the universal parent measurement whose existence follows from Sec.~\ref{sec:pairs}, and I am then left with $2^r$ measurements (some of which are parent measurements) giving rise to $(\IRG_{2,n})^r$.
Note that, whenever $l\neq0$, an asymmetry is introduced by the choice of which measurement is not paired with another one, but this problem (recall that all marginals in Eq.~\eqref{eqn:irg} should feature the same parameter $\eta$) can be overcome by symmetrisation, namely, by averaging over suitable choices of $2l$ pairs of measurements among the $k$ available.

Let me illustrate this procedure on a triplet $\{A,B,C\}$ of measurements in dimension $n$.
For any pair $\{A,B\}$ of measurements, I denote by $G(A,B)$ their parent measurement used to derive the universal lower bound in Sec.~\ref{sec:pairs}.
Importantly, the form of this parent measurement is completely explicit (as a function of the measurements $A$ and $B$) only for rank-one measurements, see Eq.~\eqref{eqn:irg_low_ansatz}, but its existence is nonetheless ensured in general in virtue of the postprocessing monotonicity of $\irg$, see Ref.~\cite{DFK19}.
Then the following measurement is a valid parent measurement in Eq.~\eqref{eqn:irg}:
\begin{equation}
  \label{eqn:mix}
  \frac13\Big[G\big(G(A,B),C\big)+G\big(G(C,A),B\big)+G\big(G(B,C),A\big)\Big],
\end{equation}
and gives a bound on $\irg_{\{A,B,C\}}$ which is general and thus applies to $\IRG_{3,n}$ defined in Eq.~\eqref{eqn:IRG}, namely,
\begin{equation}
  \IRG_{3,n}\geq\IRG_{2,n}\frac{2\IRG_{2,n}+1}{3}.
  \label{eqn:three_low}
\end{equation}
Intuitively, in each term of Eq.~\eqref{eqn:mix}, one measurement goes through $G$ once, only acquiring a factor $\IRG_{2,n}$, and the other two are fed to $G$ twice, thus getting a factor $(\IRG_{2,n})^2$.
More formally, with a slight abuse of notations, for the first term of Eq.~\eqref{eqn:mix}, I can explicitly compute the various marginals using the properties of $G$, namely, to satisfy, for all pairs of measurements in dimension $n$, the constraints in Eq.~\eqref{eqn:irg} for $\eta=\IRG_{2,n}$.
For the first marginal, the inequality
\begin{equation}
  \sum_{ab}G\big(G(A_a,B_b),C_c\big)\geq\IRG_{2,n}C_c
\end{equation}
comes directly since the indices $a$ and $b$ label all outcomes of $G(A,B)$.
For the second one, I get
\begin{align}
  \sum_{ac}G\big(G(A_a,B_b),C_c\big)&\geq\sum_a\IRG_{2,n}G(A_a,B_b)\\
  &\geq(\IRG_{2,n})^2B_b
\end{align}
and similarly for the last one, that is, the sum over $b$ and $c$.
Then Eq.~\eqref{eqn:three_low} arises from the average of the three symmetrised terms in Eq.~\eqref{eqn:mix}.

In general, for $k=2^r+l$ with $0\leq l<2^r$, for one term of the symmetrised parent measurement, there are $k-2l=2^{r+1}-k$ terms that undergo the pairwise parent measurement $r$ times and $2l=2(k-2^r)$ that undergo it $r+1$ times.
As the symmetrisation takes care of evening this noise among all $k$ measurements, I end up with a general lower bound as a function of the pairwise one, namely,
\begin{equation}
  \IRG_{k,n}\geq\left(\IRG_{2,n}\right)^r\left[1-2\,\Big(1-\IRG_{2,n}\Big)\left(1-\frac{2^r}{k}\right)\right],
  \label{eqn:more_low}
\end{equation}
where $r=\lfloor\log_2k\rfloor$.

This procedure often improves on Eq.~\eqref{eqn:irg_cloning}, see the bold numbers in Table~\ref{tab:low_more} for low dimensions.
Actually, the pattern observed there, namely, that the improvement occurs only for $k$ not too large, is general.
For instance, in dimension $n=100$, this happens for $k\leq31$.
Seen differently, this means that for all $k$ the bound~\eqref{eqn:more_low} is always improving on Eq.~\eqref{eqn:irg_cloning} for sufficiently large $n$.

In Ref.~\cite[Appendix~E.4]{DFK19}, I could also come up with the exact value
\begin{equation}
  \IRG_{3,2}=\frac12\left(1+\frac{1}{\sqrt{3}}\right),
  \label{eqn:qubit-triplets}
\end{equation}
reached by projective measurements onto states whose Bloch vectors form an orthonormal basis, typically, $X$, $Y$, and $Z$~\cite{Bus86}.

\section{Experimental certification}
\label{sec:certification}

\subsection{Estimation of the steering robustness}
\label{sec:sr}

The steering robustness was introduced in Def.~\ref{def:sr} in Sec.~\ref{sec:incompatibility} but it was quite abstract.
In this section I recall its interpretation as the amount of violation of an optimal steering inequality.
This approach is very useful for experiments because it allows me to estimate the steering robustness without requiring a full tomography.
The technical details presented below only provide a brief explanation for the estimation procedure given in Eq.~\eqref{eqn:estimation}.

The definition of the steering robustness in Eq.~\eqref{eqn:SR} is a SDP; this means that it comes with a dual problem~\cite{BV04}.
This dual is, for instance, stated in Ref.~\cite{UKS+19} and reads
\begin{align}\label{eqn:dual}
  \SR_{\{\sigma_{a|x}\}}=\max_{\{F_{a|x}\}}&\quad\sum_{a,x}\Tr(F_{a|x}\sigma_{a|x})-1\\
  \mathrm{s.t.}&\quad\sum_{a,x}\Tr(F_{a|x}\tau_{a|x})\leq 1\nonumber\\
  &\quad F_{a|x}\geq0\quad\forall\,a,x,\nonumber
\end{align}
where the first inequality (i)~should hold for all assemblages $\{\tau_{a|x}\}$ admitting a LHS as defined in Eq.~\eqref{eqn:LHS} and (ii)~can actually be cast into SDP constraints.
From this, the steering robustness can be interpreted as the maximal violation of all steering inequalities (defined below) normalised to have a LHS bound of 1.

Similarly to entanglement witnesses, which are hyperplanes separating the set of states into (i)~a region where all separable states (and many entangled ones) lie and (ii)~a region where entanglement can be certified, there are steering witnesses, often called steering inequalities, which amount to choosing the measurements on Bob's side in order to compute a functional whose value is bounded for unsteerable assemblages, thus whose violation demonstrates steering.
More formally, if Bob is measuring $\{B_{a|x}\}$, by choosing $F_{a|x}=B_{a|x}/\lambda$, where $\lambda$ is such that the first constraint in Eq.~\eqref{eqn:dual} is satisfied, then I get, since $F_{a|x}$ satisfies all the constraints of Eq.~\eqref{eqn:dual},
\begin{align}
  \SR_{\{\sigma_{a|x}\}}&\geq\sum_{a,x}\Tr(F_{a|x}\sigma_{a|x})-1\\
  &\geq\frac{1}{\lambda}\sum_{a,x}\Tr\Big\{B_{a|x}\Tr_A\big[(A_{a|x}\otimes\id_B)\rho_{AB}\big]\Big\}-1\nonumber\\
  &\geq\frac{1}{\lambda}\sum_{a,x}\Tr\big[(A_{a|x}\otimes B_{a|x})\rho_{AB}\big]-1,\label{eqn:estimation}
\end{align}
this last quantity being measurable experimentally since it involves the coincidence counts between Alice and Bob.

\begin{table}[ht]
  \centering
  \begin{tabular}{|c|ccccc|}
    \hline
    \diagbox{$~k~$}{$~n~$} & 2                       & 3            & 4            & 5            & 6            \\ \hline
    2                      & \bf ~0.1716~            & \bf 0.2679   & \bf 0.3333   & \bf 0.3820   & \bf 0.4202   \\
    3                      & \bf 0.2679$^\ast\!\!\!$ & \bf ~0.4759~ & \bf 0.6      & \bf ~0.6941~ & \bf ~0.7692~ \\
    4                      & 0.3333                  & 0.6          & \bf ~0.7778~ & \bf 0.9098   & \bf 1.0170   \\
    5                      & 0.3636                  & 0.6667       & 0.9231       & 1.1429       & \bf 1.2877   \\
    6                      & 0.3846                  & 0.7143       & 1            & 1.25         & 1.4706       \\
    7                      & 0.4                     & 0.75         & 1.0588       & 1.3333       & 1.5790       \\
    8                      & 0.4118                  & 0.7778       & 1.1053       & 1.4          & 1.6667       \\ \hline
  \end{tabular}
  \caption{
    Upper bounds on the steering robustness $\SR$ obtained by combining Eq.~\eqref{eqn:SRup2} with the results of Sec.~\ref{sec:universal}.
    The roman values are obtained with the cloning machine bound in Eq.~\eqref{eqn:irg_cloning}.
    The bold ones indicate the cases for which I could improve on this bound: the star refers to Eq.~\eqref{eqn:qubit-triplets} while the remaining bold values follow from the method elaborating on the bound for pairs of measurements, resulting in Eq.~\eqref{eqn:more_low}.
  }
  \label{tab:low_more}
\end{table}

\begin{table}[hb]
  \scalebox{0.98}{
    \begin{tabular}{|c|cccccc|}
      \hline
      \backslashbox{$~k~$}{$~d~$} & 2          & 3          & 4          & 5             & 6          & 7                   \\ \hline
      2                           & $~0.1716~$ & $~0.2679~$ & $~0.3333~$ & $~0.3820~$    & $~0.4202~$ & $~0.4514~$          \\
      3                           & $0.2679$   & $0.4037$   & $0.5$      & $0.6001^\ast\!\!\!$ & $0.6544$   & $0.7106$      \\
      4                           &            & $0.5279$   & $0.6$      & $0.7568$      &            & $0.8955^\ast\!\!\!$ \\
      5                           &            &            & $0.7446$   & $0.8716$      &            & $1.0161$            \\
      6                           &            &            &            & $0.9645$      &            & $1.1085$            \\
      7                           &            &            &            &               &            & $1.1801$            \\
      8                           &            &            &            &               &            & $1.3407$            \\\hline
    \end{tabular}
  }
  \caption{
    Value of the steering robustness $\SR$ theoretically reachable with projective measurements onto $k$ MUBs in dimension $d$ and a maximally entangled shared state.
    The stars indicate the existence of unitarily inequivalent subsets of MUBs reaching different robustnesses~\cite{DSFB19}, in which case I only give the highest one.
    When the value of $\SR$ at given $k$, $d$, and $n$ is strictly smaller than the bound on $\IRG_{k,n}$ from Table~\ref{tab:low_more}, this means that $n$-preparability can be ruled out, i.e., that genuine $(n+1)$-dimensional steering can be demonstrated.
    As an example, a perfect setup featuring $k=5$ MUBs in dimension $d=5$ would lead to a demonstration of genuine four-dimensional steering since $0.8716>0.6667$.
  }
  \label{tab:MUB}
\end{table}

\subsection{Dimension certificate}
\label{sec:certificate}

I start by recalling the result from Ref.~\cite{DSU+21} to highlight the general mechanism of the method, before explaining how the technical results on incompatibility obtained in Sec.~\ref{sec:universal} can be used.
For pairs of measurements, that is, $k=2$, by combining Eqs.~\eqref{eqn:SRup2} and \eqref{eqn:IRG2} I have that an $n$-preparable assemblage $\{\sigma_{a|x}\}$ must satisfy
\begin{equation}\label{eqn:most-incompatible}
  \SR_{\{\sigma_{a|x}\}}\leq\frac{\sqrt{n}-1}{\sqrt{n}+1},
\end{equation}
or equivalently
\begin{equation}\label{eqn:ceiling}
  n\geq\left(\frac{1+\SR_{\{\sigma_{a|x}\}}}{1-\SR_{\{\sigma_{a|x}\}}}\right)^2.
\end{equation}
Violating this inequality thus rules out all $n$-preparable assemblages, demonstrates genuine high-dimensional steering, and provides at the same time a one-sided device-independent dimension witness for the underlying state.
Since the right-hand side of Eq.~\eqref{eqn:ceiling} is an increasing function of $\SR_{\{\sigma_{a|x}\}}$, it is possible to plug any lower bound on its value to get a valid bound for $n$, whose violation will again give the desired behaviour.
Therefore, the bounds from Sec.~\ref{sec:sr} can be used.

In a nutshell, the procedure consists in the following steps: choosing the steering inequality depending on the state and measurements used, measuring the amount of violation experimentally obtained to get a lower bound on the steering robustness, and finally use this estimate to exclude low dimensions and thus show genuine high-dimensional steering.

When the number of measurements is strictly bigger than two, the most incompatible sets of measurements in a given dimension are not known.
However, the bounds presented in Sec.~\ref{sec:more} can still be applied to get dimension witnesses.
There are admittedly not tight but will nonetheless turn useful.

First, the cloning machine bound~\eqref{eqn:irg_cloning} already gives an answer: for an $n$-preparable assemblage $\{\sigma_{a|x}\}$, the inequality
\begin{equation}
  n\geq 1+\frac{2k\,\SR_{\{\sigma_{a|x}\}}}{k-\SR_{\{\sigma_{a|x}\}}-1}
  \label{eqn:dim_witness_cloning}
\end{equation}
holds.
Note that this bound already gives nontrivial certificates.

Second, I can take advantage of my better bound~\eqref{eqn:more_low} (from a certain dimension on) to derive an improved witness.
Given the form of Eq.~\eqref{eqn:more_low}, an analytical solution valid for all $k$ cannot be found.
Restricting to $k=2^r$, I can nonetheless derive the witness
\begin{equation}
  n\geq\left(\frac{\left(1+\SR_{\{\sigma_{a|x}\}}\right)^{\frac1r}}{2-\left(1+\SR_{\{\sigma_{a|x}\}}\right)^{\frac1r}}\right)^2,
  \label{eqn:dim_witness_power_two}
\end{equation}
which indeed recovers Eq.~\eqref{eqn:ceiling} when $r=1$.
The other values of $k$ can be solved case by case, preferably numerically.
Nonetheless, when $k=3$, I get the explicit witness
\begin{equation}
  n\geq\frac{\left(1+\SR\right)\left(17+5\,\SR+\sqrt{\left(1+\SR\right)\left(25+\SR\right)}\right)}{8\left(2-\SR\right)^2},
  \label{eqn:dim_witness_three}
\end{equation}
where I have omitted the indices $\{\sigma_{a|x}\}$ for conciseness.
Moreover, the minimal value of $\SR_{\{\sigma_{a|x}\}}$ needed to demonstrate genuine three-dimensional steering can be further lowered from $(53-36\sqrt2)/7\approx0.2983$ predicted by Eq.~\eqref{eqn:dim_witness_three} to $(3-\sqrt3)/(3+\sqrt3)\approx0.2679$ obtained by using the tight lower bound~\eqref{eqn:qubit-triplets}.

\begin{figure*}[ht!]
  \centering
  \includegraphics[height=0.33\textwidth]{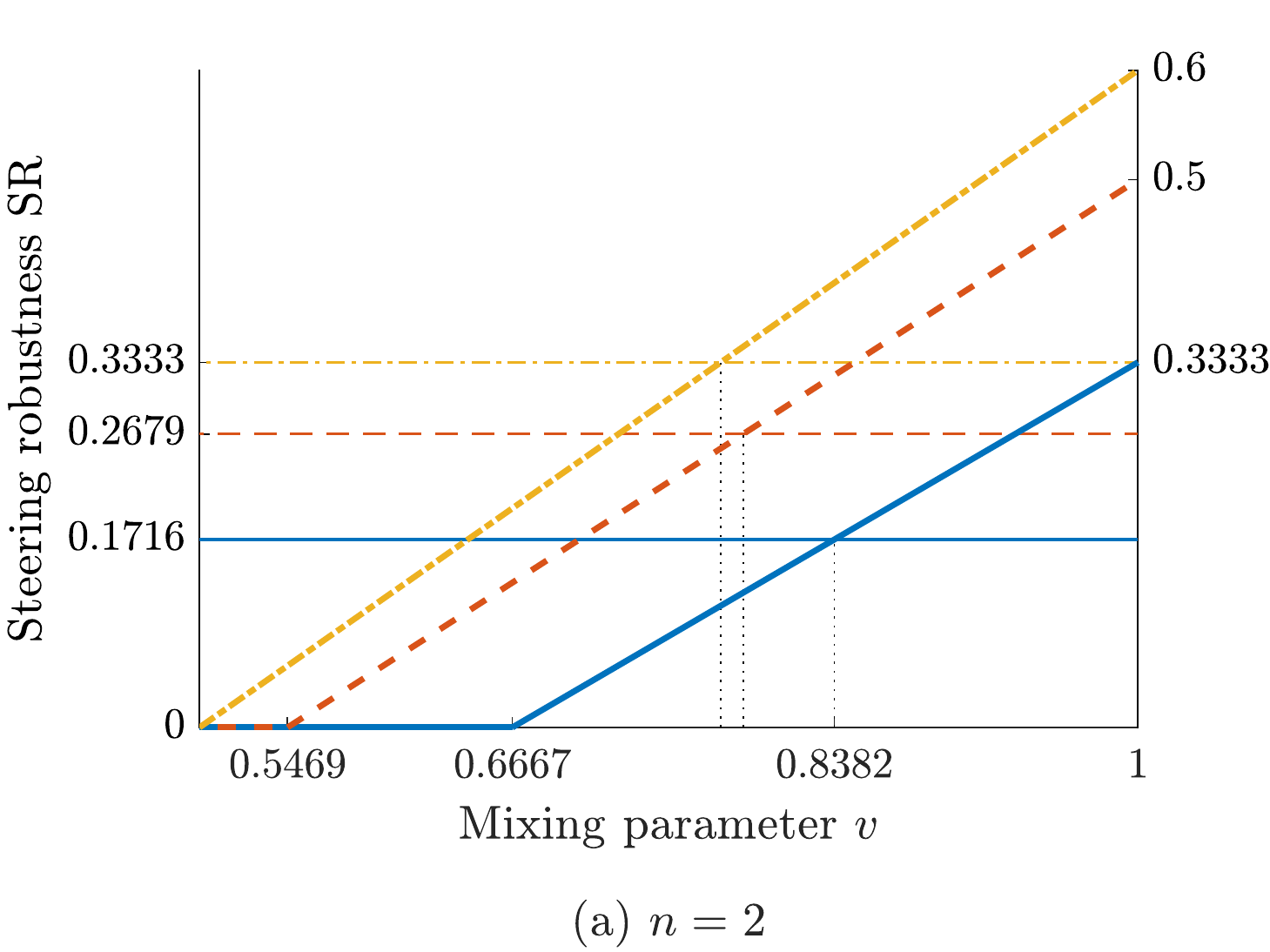}
  \hspace{0.01\textwidth}
  \includegraphics[height=0.33\textwidth]{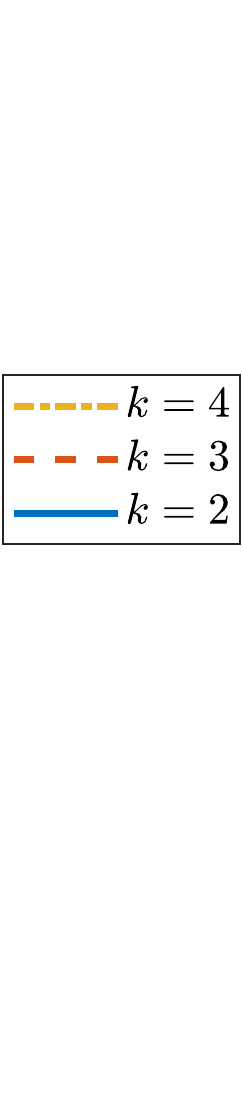}
  \hspace{0.01\textwidth}
  \includegraphics[height=0.33\textwidth]{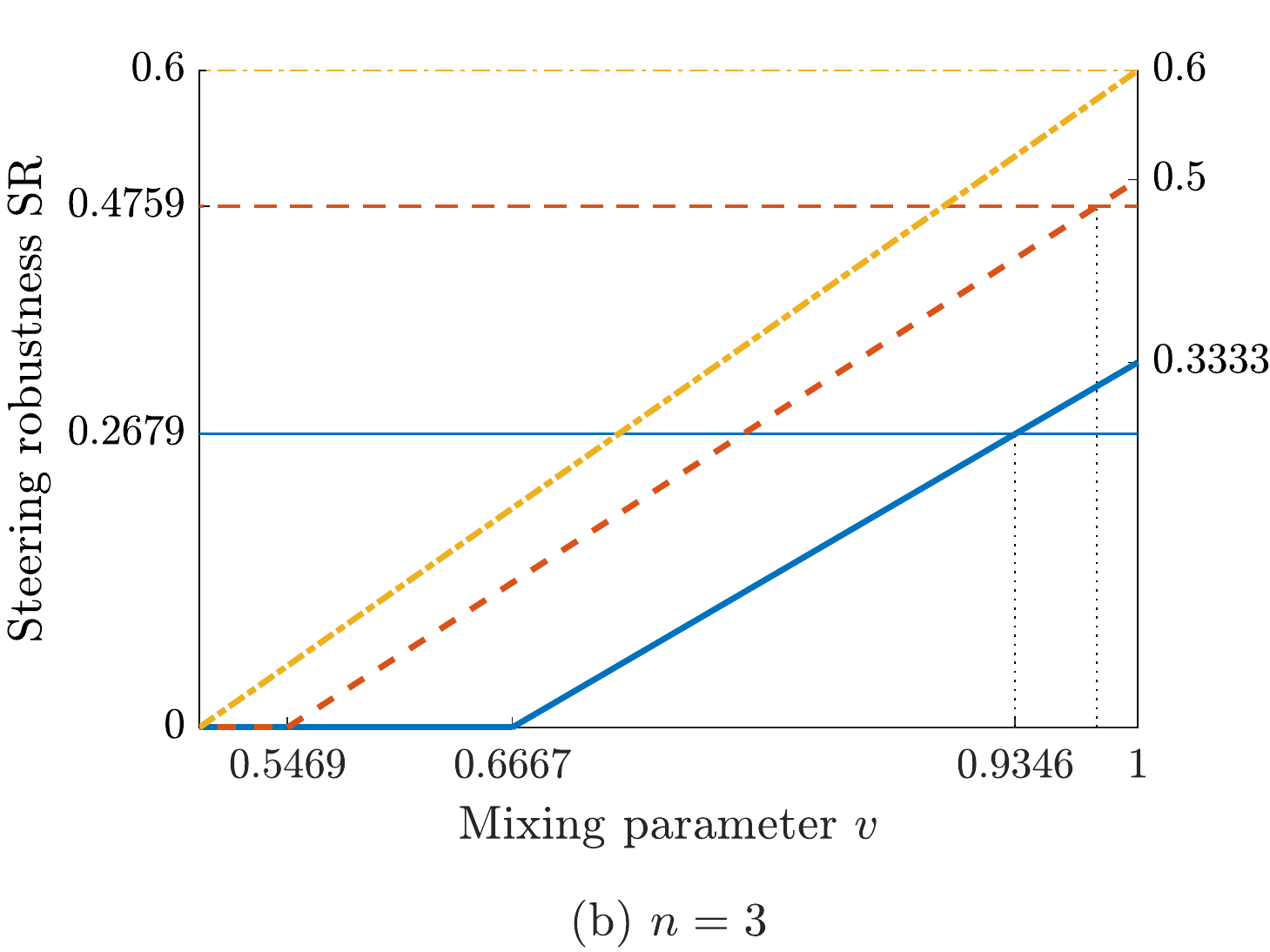}
  \caption{
    Effect of using a noisy maximally entangled state with mixing parameter $v$, see Eq.~\eqref{eqn:iso}, on the steering robustness achievable in dimension $d=4$ with projective measurements onto $k=2,3,4$ MUBs.
    The horizontal lines correspond to the bounds given in Table~\ref{tab:low_more}.
    The crossings on the horizontal axis are predicted by Ref.~\cite{DSFB19} while those on the vertical axis (on the right) are simply given in Table~\ref{tab:MUB}.
    In the example chosen, using more than two measurements yields an improved resistance to noise for $n=2$ but not $n=3$.
    Note that the degeneracies happening at $\frac13$ and $\frac35$ are coincidental.
  }
  \label{fig:sr}
\end{figure*}

\section{Example}
\label{sec:example}

In this section, I give an example for the reader interested in conducting an experiment with concrete measurements.
Indeed, even though the best measurements to demonstrate genuine high-dimensional steering are not known in general, there is a fairly natural choice that performs quite well.
Moreover, with the certificates developed above, the resistance to noise of the dimension that can be demonstrated using these measurements often increases with their number.

\subsection{State and measurements used}
\label{sec:MUB}

In the following, the shared state will always be of the form
\begin{equation}
  \label{eqn:iso}
  \rho_{AB}(v):= v\ketbra{\phi_d} + (1-v)\frac{\id_{AB}}{d^2},
\end{equation}
where ${\ket{\phi_d}:=\sum_{i=0}^{d-1}\ket{ii}/\sqrt{d}}$ is the maximally entangled state and $v$ is the mixing parameter.
Then, the longer genuine high-dimensional steering survives when decreasing $v$, the more robust it is.

Let me now introduce mutually unbiased bases (MUBs).
These orthonormal bases in dimension $d$ are such that, when taking any two of them, denoted $\{\varphi_i\}_i$ and $\{\psi_j\}_j$, the scalar product $|\braket{\varphi_i}{\psi_j}|$ is always equal to $1/\sqrt{d}$ for all $i$ and $j$.
They generalise $X$, $Y$, $Z$ in the Bloch sphere and are ubiquitous in quantum information (see, e.g., Ref.~\cite{DEBZ10}).
Note that there cannot be more than $d+1$ MUBs in dimension $d$ and that a construction of a complete set of $d+1$ MUBs is only known when $d$ is a power of a prime number; from now on I will always refer to the construction in Ref.~\cite{WF89} when discussing MUBs.
Already in Ref.~\cite{DSU+21} the measurements implemented were projective measurements onto pairs of MUBs.
For more than two measurements, that is, $k>2$, though subsets of $k$ MUBs are generally not optimal (meaning that there are other sets of $k$ measurements able to reach a higher steering robustness, see Ref.~\cite{BQG+17}), they still have reasonable robustnesses as can be seen in Table~\ref{tab:MUB}, which should be compared with Table~\ref{tab:low_more} to understand how the certificates work.

For projective measurements onto MUBs, a good choice of $\{B_{a|x}\}$ defined in Sec.~\ref{sec:sr} is given by ${B_{a|x}=A_{a|x}^T}$, that is, the transposed of the measurements performed on Alice's side.
For $d\leq8$, the only case for which this is not optimal is when $k=4$ and $d=5$, see Ref.~\cite[Table~2.2]{PhD}.
In general, a proof of optimality is only known for $k=2$, $k=d$, and $k=d+1$~\cite{ULMH16,DSFB19,NDBG20}, the last two cases being restricted to the standard construction of MUBs~\cite{WF89}.

\subsection{Resistance to noise}
\label{sec:resistance}

I can now state a good property that the bounds derived in Sec.~\ref{sec:universal} enjoy even though they are not optimal: the resulting dimension certificates often show a higher resistance to noise than the one for pairs of measurements.
More precisely, see Fig.~\ref{fig:sr} where the steering robustness is plotted as a function of the mixing parameter $v$ from Eq.~\eqref{eqn:iso}.
The fact that the resulting curve is a line follows from the linearity of the underlying equations.
The value of $v$ for which $\SR$ starts to be positive is exactly the one predicted in Ref.~\cite{DSFB19} while the value of $\SR$ for $v=1$ is given in Table~\ref{tab:MUB}.
The natural advantage of increasing the number $k$ of measurements (i.e., the diagonal lines in Fig.~\ref{fig:sr} go up) is then counterbalanced by the looseness of the dimension witnesses derived above (i.e., the horizontal lines in Fig.~\ref{fig:sr} also go up, but faster).
In many cases, the overall effect still gives an advantage over using $k=2$ measurements, see Table~\ref{tab:noise}.
Note that the threshold in this last case was derived in Ref.~\cite{DSU+21} and reads
\begin{equation}
  \frac{\left(d+\sqrt{d}-1\right)\sqrt{n}-1}{(d-1)\left(\sqrt{n}+1\right)},
\end{equation}
while an analytical value for the other cases is less elegant as it would typically involve the optimal eigenvalues defined in Ref.~\cite{DSFB19}.

\begin{table*}[ht!]
  \begin{minipage}{0.48\textwidth}
    \centering
    \begin{tabular}{|c|ccccc|}
      \hline
      \diagbox{$~n~$}{$~d~$} & 3        & 4          & 5          & 6          & 7          \\ \hline
      2                      & ~0.8860~ & ~~0.8382~~ & ~~0.8097~~ & ~~0.7899~~ & ~~0.7751~~ \\
      3                      &          & 0.9346     & 0.8969     & 0.8713     & 0.8525     \\
      4                      &          &            & 0.9560     & 0.9266     & 0.9051     \\
      5                      &          &            &            & 0.9677     & 0.9442     \\
      6                      &          &            &            &            & 0.9749     \\ \hline
    \end{tabular}
    \\[5pt]
    (a) $k=2$
  \end{minipage}
  \begin{minipage}{0.48\textwidth}
    \centering
    \begin{tabular}{|c|ccccc|}
      \hline
      \diagbox{$~n~$}{$~d~$} & 3           & 4           & 5           & 6           & 7           \\ \hline
      2                      & \bf~0.8549~ & \bf~0.78977 & \bf~0.7405~ & \bf~0.7168~ & \bf~0.6948~ \\
      3                      &             & 0.9781      & 0.9030      & \bf0.8692   & \bf0.8382   \\
      4                      &             &             & 0.99992     & 0.9602      & 0.9238      \\
      5                      &             &             &             &             & 0.9887      \\ \hline
    \end{tabular}
    \\[5pt]
    (b) $k=3$
  \end{minipage}
  \\[10pt]
  \begin{minipage}{0.48\textwidth}
    \centering
    \begin{tabular}{|c|ccccc|}
      \hline
      \diagbox{$~n~$}{$~d~$} & 3           & 4           & 5           & 7           & 8           \\ \hline
      2                      & \bf~0.8090~ & \bf~0.7778~ & \bf~0.6987~ & \bf~0.6507~ & \bf~0.6210~ \\
      3                      &             &             & \bf0.8884   & \bf0.8164   & \bf0.7742   \\
      4                      &             &             &             & 0.9269      & \bf0.8763   \\
      5                      &             &             &             &             & 0.9521      \\ \hline
    \end{tabular}
    \\[5pt]
    (c) $k=4$
  \end{minipage}
  \begin{minipage}{0.48\textwidth}
    \centering
    \begin{tabular}{|c|ccccc|}
      \hline
      \diagbox{$~n~$}{$~d~$} & 4           & 5           & 7           & 8           & 9           \\ \hline
      2                      & \bf~0.7089~ & \bf~0.6608~ & \bf~0.6179~ & \bf~0.5824~ & \bf~0.5643~ \\
      3                      & 0.9405      & \bf0.8631   & \bf0.7953   & \bf0.7439   & \bf0.7175   \\
      4                      &             &             & 0.9455      & \bf0.8805   & \bf0.8471   \\
      5                      &             &             &             & 0.9976      & 0.9581      \\ \hline
    \end{tabular}
    \\[5pt]
    (d) $k=5$
  \end{minipage}
  \caption{
    Noise thresholds of the dimension certificates when using projective measurements onto $k$ MUBs and the shared state~\eqref{eqn:iso}.
    The bold values indicate the instances for which, when $k>2$, the resistance is improved with respect to the case $k=2$.
    Empty or nonexistent cells mean that no genuine $(n+1)$-dimensional steering could be shown, even with a perfect state, i.e., when $v=1$ in Eq.~\eqref{eqn:iso}.
  }
  \label{tab:noise}
\end{table*}

\section{Conclusion}
\label{sec:conclusion}

In this paper, I have recalled how to experimentally demonstrate genuine high-dimensional steering: using a steering inequality to bound the steering robustness and exploiting its violation to witness the high-dimensional nature of the underlying state, and this without trusting one of the parties.
Going beyond the limitation of the previous work on the topic, namely, being restricted to pairs of measurements, I have reviewed and derived general incompatibility bounds for many measurements that could be turned into dimension certificates.
Although these witnesses are not tight, they nonetheless often exhibit the desirable feature of tolerating more noise than the ones obtained from pairs of measurements.
This may be relevant for experiments because it provides a more robust way of demonstrating genuine high-dimensional steering.
Note that this task was not done in the recent experimental works of Refs.~\cite{LLCY18,WPD+18,ZWLZ18,QWA+21} since they demonstrate standard steering and not genuine high-dimensional steering that would give a dimension certificate as a byproduct.

The example used to illustrate the procedure relies on projective measurements onto MUBs and a noisy maximally entangled state.
This does not mean that the method is restricted to these; any sufficiently incompatible measurements may indeed give rise to similar results and should be discussed case by case.

Investigating whether the resistance to loss, that is, when the untrusted party uses imperfect detectors, increases or not with the current bounds could also reveal an advantage of using more than two measurements.
Moreover, thanks to the connection between the problem of certifying dimension in a one-sided device-independent way and the most incompatible sets of measurements, any advances on this latter side would immediately have consequences on the former.
Hopefully this will encourage further theoretical work in the incompatibility community.

\section*{Acknowledgements}

The author thanks Nicolas Brunner and Roope Uola for discussions.
Financial support from the Swiss National Science Foundation (Starting grant DIAQ and NCCR QSIT) is gratefully acknowledged.

\newpage

\bibliography{main}
\bibliographystyle{sd2}

\end{document}